\documentclass{ws-procs10x7}

\makeatletter
\newcommand{\subsubsection}%
        {\@startsection{subsubsection}{3}{0em}%
        {-3.25ex\@plus -1ex \@minus -.2ex}%
        {1.5ex \@plus .2ex}%
        {\normalfont\itshape}}
\makeatother

\begin{document}


%

%



\title{
Searching for new physics
 at future accelerators
}

\author{Riccardo Barbieri}

\address{Scuola Normale Superiore, Pisa, and INFN, Italia } 


\twocolumn[\maketitle\abstract{
I overview the status of the Electroweak Symmetry Breaking problem, paying special attention to the possible signals of new physics at the Large Hadron Collider (and at a Linear Collider)}]


\newpage

\setcounter{page}{1}
\setcounter{footnote}{0}

\section{Introduction and a Calculability Principle}
\label{int}

To talk of the future, it is both useful and reassuring to start by recalling the great synthesis of the Standard Model, encapsulated in the three, or four, little lines of its Lagrangian
\begin{eqnarray}
\mathcal{L}=&-&\frac{1}{4}  F^a_{\mu\nu} F^{a\, \mu\nu}
+ i \bar \psi  \not\!D \psi  \\
&+& \psi^T \lambda \psi h + h.c.  \\
&+& |D_{\mu} h|^2 - V(h)  \\
&+&  N^T M N  
\end{eqnarray}
The notation, concise but precise as well, is self-explanatory. Flavor, gauge  and spinor indices are left understood. To account for the neutrino masses,   $\psi$, a vector of Weyl spinors describing all matter, includes  the right handed neutrino, $N$, one per generation, which may be heavy and decoupled - hence the mass terms in the fourth line - if the observed left-handed neutrinos are Majorana. By 
including the neutrino masses, I am using a definition of the SM which is broader than the historical one. I find this appropriate and, I hope, not confusing.

Each line of this Lagrangian corresponds to a different sector of the theory. The gauge sector (line 1) is by far the best tested one. The flavor \cite{AGLP}  and the neutrino-mass \cite{LMW} sectors (lines 2 and 4) have witnessed the main developments in the last years, of different nature but both of the greatest significance. I find the Electroweak Symmetry-Breaking  (EWSB) sector (line 3) as still the most elusive and the one that is more likely to deserve surprises. It is certainly true that even this sector has passed, so far, all the Electroweak Precision Tests (EWPT)\cite{LEPEWWG}, now crucially including also the recently published LEP2 results \cite{ALEPH} \footnote{ See ref. (\cite{BPRS}) for an analysis that underlines the significance of LEP2 to test "universal" theories of EWSB with a mass gap.}. One cannot forget, nevertheless, the indirect nature  of these tests, to be contrasted with the scanty direct experimentation at energies well above the Fermi scale, $G_F^{-1/2}$. Therefore it is on the EWSB sector that I concentrate my attention in the following.

The EWSB problem has received and still receives indeed a lot of theoretical attention, with many different proposals to solve or at least to address it. All  such proposals involve a new physical scale, $\Lambda_{NP}$. To constrain the field of the discussion, I limit my attention to the proposals that satisfy a "Calculability Principle": the Z mass, or the Fermi scale, (and the Higgs mass, when a Higgs boson exists), should be related in a calculable manner to $\Lambda_{NP}$, so that

\begin{equation}
\label{cal-eq}
M_Z =  \Lambda_{NP} f ( a_i),
\end{equation}
where $f(a_i)$ is a calculable function of the physical parameters $a_i$. An  example that satisfies this requirement is technicolor\cite{SW}, where

\begin{equation}
M_Z =  \frac{g_2}{2 \cos{\theta_W}} f_{\pi_{TC}},
\end{equation}
and $ f_{\pi_{TC}}$ is the techni-pion decay constant. As a further requirement, at $\Lambda_{NP}$ the theory should be sufficiently under control that its consistency with the EWPT can be explicitly checked. This view can admittedly be a bit too narrow. Note also that, since some of the parameters  $a_i$ will in general be unknown, eq. (\ref{cal-eq})  alone may not be sufficient to constrain $\Lambda_{NP}$ in a significant manner. Nevertheless, since I want to make the discussion concrete, I prefer to stick to these requirements. If I do, all currently acceptable proposals for EWSB involve, as far as I can tell and after some little further qualifications, either supersymmetry or the Higgs as a Pseudo-Goldstone-Boson (PGB) associated with the breaking of some appropriate global symmetry. I therefore discuss them in turn.

\section{Supersymmetry}\label{SUSY}
\subsection{ The "problem" of the Minimal Supersymmetric Standard Model}

The merits of Supersymmetry, as realized in the Minimal Supersymmetric Standard Model (MSSM)\cite{MSSM} \cite{unif-mod}, are well known: 
\begin{itemize}
\item It can explain the unification of the gauge couplings\cite{unif}
\item It predicts the Higgs mass
\item It smoothly passes the EWPT
\item It provides a natural dark matter candidate in the form of a neutralino.
\end{itemize}
This explains the large interest in the MSSM and the focus on it even of the experimental searches at the future accelerators. The crucial question is therefore: Where are the superpartners? If the theory is true, will they have to be seen at the Large Hadron Collider (LHC)?

The only way we have to address these questions at present is to consider eq. (\ref{cal-eq}) in the specific version that it takes in the MSSM
\begin{equation}
\label{MZ^2}
M_Z^2 \approx  (90 GeV)^2 (\frac{< m_{\tilde{t}}>}{230 GeV})^2 
\log (\frac{\Lambda_{UV}}{ < m_{\tilde{t}}>}) + \ldots,
\end{equation}
where $< m_{\tilde{t}}>$ is a suitable average of the stop masses, $\Lambda_{UV}$ is a model-dependent ultraviolet cut-off and the dots stand for a number of other terms, of both signs, dependent on parameters not related, as far as we know, to the ones appearing in the term shown in eq. (\ref{MZ^2}).
If these terms intervene to give a near  cancellation of the one explicit in (\ref{MZ^2}), not much can be said on the s-partner masses other than they have to be compatible with the current lower limits. Since we do not see a reason for this cancellation, however, we can  bar it. If we do, then the stop and the other s-particle masses, at various degrees, get limited from above\cite{BG}. At the same time however, always with reference to eq.(\ref{MZ^2}), there is a problem\cite{CEP}: since $< m_{\tilde{t}}> $ can hardly be lighter than 500 GeV or so without suppressing the radiative corrections to the Higgs mass needed for consistency with the negative Higgs searches, the term shown in eq. (\ref{MZ^2}) gets much too large. Note that it 
grows quadratically with $< m_{\tilde{t}}>$, whereas the Higgs mass grows only logarithmically (and must be calculated including the two loop corrections\cite{all} for a proper comparison with the experimental lower bound of 115 GeV \cite{LEPEWWG}).
It should be clear why this is a "problem" in quotation marks. Let us see what I call the reactions to it.

\subsection{The reactions to the problem}
\subsubsection{Never mind a few per-cent tuning}

Numerically, the cancellation that is needed in eq. (\ref{MZ^2}) is at the few per-cent level, at least, provided the Higgs is just around the corner\cite{fine-tun}.  In nature, on the other hand, there are apparently accidental cancellations at the per-cent level, or even stronger. This could just be another one. If this is the case, it is reassuring to know that the s-particle searches at the LHC will explore much of the parameter space of the MSSM up to a tuning at the per-mil level. Will this be enough to discover, other than the Higgs, (some of) the superpartners? Difficult to say, now that we know that some fine tuning is there. The signals are by now well known, in any case.

\subsubsection{Increase $m_h^{tree}$ by an extra bit of quartic coupling}

The need of a large radiative contribution to the Higgs mass - and therefore of a large $< m_{\tilde{t}}>$ - comes about because of the MSSM bound on the tree-level Higgs mass, $m_h^{tree} \leq M_Z$. In turn, this is so because in the MSSM the quartic Higgs self-coupling at tree level is a gauge coupling. Ways to change this exist, however, and have been proposed.
To the least it requires adding a gauge singlet supermultiplet with a Yukawa coupling to the Higgs. This defines the so called NMSSM\cite{NMSSM}. Or it may be obtained by enlarging the gauge group\cite{extra-gauge}. In either cases the thing to watch is gauge unification, which may be affected by these changes. For example in the NMSSM, this severely limits the amount by which one can raise the Higgs mass. Recent proposals\cite{heavier-Higgs} get around this limitation by adding extra structure, so that the Higgs mass can be significantly raised, even well above the upper bound of about 250 GeV at $95 \%$ C.L. from the EWPT in the SM\cite{LEPEWWG}. This is technically possible because of extra parameters that affect the EWPT other than the Higgs mass itself\cite{HB}. Technically possible yes, but plausible? In any event the phenomenology of these extended models may differ from the one of the MSSM in various instances, especially related to the Higgs sector.

\subsubsection{Reduce the tuning in $M_Z =  \Lambda_{Susy} f ( a_i)$ }

Modifications of the MSSM that reduce the fine tuning in eq. (\ref{MZ^2}) are being looked for extensively, with several ingenious suggestions\cite{reduce-tun}. In principle, my preference goes in the direction of looking for alternative ways to break supersymmetry, that could have an impact also on the fine-tuning problem. The example that I like most uses the boundary conditions on a fifth dimension of length $\pi R/2$ to break supersymmetry\cite{S-S} \cite{BHN}. 
Clear consistency with the EWPT\cite{PM} requires a partial localization in the fifth dimension of the top supermultiplet, which allows to raise $1/R$ in the range from 1.5 to 4 TeV while still maintaining  the absence of fine tuning\cite{loc-top}. A non negligible merit of this model is that only one Higgs scalar gets a non-zero vacuum expectation value consistently with supersymmetry and with the various phenomenological requirements\cite{BHN}. Most of the s-partners are heavy with a mass at $1/R$. The Higgs mass is just above the current bound by $10 \div 15$ GeV at most. The striking signature is a stable or quasi-stable third-generation s-fermion\cite{BHN}, which becomes the LSP and has a mass, proportional to $1/R$, in the $500\div1300$ GeV\cite{two-loops}. The price to pay, however, is that one gives up the conventional unification of couplings and the dark matter neutralino.

\subsubsection{The absence of tuning is not the criterium at all }

A more drastic departure from the mainstream has been recently put forward\cite{split-susy}, inspired by the frustration
about the cosmological constant problem and by recent string theory developments. Maybe supersymmetry is there, but its role is not to protect the Higgs mass, or the Higgs vev, from large radiative corrections. Both the Higgs mass and its vev, or the Z mass,  are highly fine tuned, since all the sfermions, including the stops, are very heavy, outside the reach of the colliders, present and future. What is maybe at reach of the colliders  are the gauginos and the higgsinos, as suggested by the gauge coupling unification and by the interpretation of the dark matter as a stable neutralino. Note that this "Split Supersymmetry" is nothing but  the MSSM itself in a particular region of its parameter space, - generally not considered because highly fine tuned -, which has a striking signature: a stable or quasi stable gluino\cite{split-susy}.

\section{Strongly interacting models of ESWB}

In the past few years there has been a revival of interest in strongly interacting models of EWSB. To some extent this is surprising, since we used to think that, when they are calculable, they do not work with the EWPT. Standard technicolor is the example of this. The reason of this revival has largely to do with the model building in extra-dimensional theories. Paradoxically, in some cases, the introduction of an extra dimension leads to an improvement of the calculability. Often one can view these models both as concrete extra-dimensional physical theories or  as a tool to study particular strongly interacting theories in 4D. Ideas like "deconstruction" \cite{dec} or the AdS/CFT correspondence\cite{AdS/CFT} are instrumental to this dual aspect. I do not intend to overview here all the different proposals: composite Higgs theories\cite{ADMS}, Higgsless theories\cite{Higgsless} \cite{BPRS},   etc., but I limit myself to discuss  the physical picture where the Higgs is a PGB of an appropriate global symmetry. It is concrete enough, so that one can say that it satisfies the Calculability Principle and that it passes the EWPT.
One may be surprised by seeing it quoted in the context of extra-dimensional theories: after all, the idea is around since quite some time\cite{GK} with no reference at all to extra dimensions. The model building has received, however, some new momentum from extra dimensions\cite{PGB-in-5D}, which in turn have inspired a particular 4D realization of the PGB picture in Little Higgs (LH) models\cite{LHM} \cite{SHM} \cite{Sc}. 

\subsection{Little Higgs models}

To realize the Higgs as a PGB, one postulates a dynamics with a global symmetry group $G_{gl}$ spontaneously broken to a subgroup $H_{gl}$ at a mass scale $f$. Part of the global symmetry group, $G_l$ is gauged.  As a result of the breaking $G_{gl} \rightarrow H_{gl}$ also the gauge group is broken to $H_l$. Two elements are essential in this construction. The unbroken gauge group $H_l$ contains the SM gauge group. Among the uneaten Goldstone bosons  there is at least one with the usual Higgs quantum numbers under the SM gauge group.

If this is the basis of any construction, it is also far from the end. The Goldstone boson of the strong dynamics which is interpreted as the Higgs must have also a self coupling, a Yukawa coupling to the fermions and a negative squared mass that destabilizes it and leads to electromagnetism as the only residual gauge group. None of these properties is trivial to achieve in an overall consistent way and insisting on a perturbatively calculable picture of EWSB without excessive fine tunings. Little Higgs models introduce suitable (somewhat {\it ad hoc}) tricks designed to solve these problems. 

There are (too) many LH models in the literature. Of special interest are the "littlest" one\cite{LHM} and the "simplest" one\cite{SHM} \cite{Sc}, with the names used by the authors themselves. The first is characterized by $G_{gl} = SU(5)$, $H_{gl} = SO(5)$ and $G_{l} = (SU(2)XU(1))^2$; the second by $G_{gl} = (SU(3)XU(1))^2$, $H_{gl} = (SU(2)XU(1))^2$ and $G_{l} = SU(3)XU(1)$. The repeated group factors are indirectly reminiscent of the extra dimensions.

A relatively model independent signature of LH models is the existence of a heavy top-like quark, $T$, or maybe a series of them, singly produced in a hadron collider via $q b \rightarrow q' T$, which then decays as $T \rightarrow t h, t Z, b W$. A LH model also includes, for sure, heavy extra gauge bosons coupled to the fermions, however, in a model dependent way. The chances of finding some of these signatures at the LHC\cite{PPP} crucially depend on the value of the breaking scale $f$, which can be called $\Lambda_{LH}$. In turn this scale can again be bounded from above only by barring accidental cancellations in the Higgs mass, or in the Fermi scale. In this respect, I find that it is not easy to do relatively better than in the MSSM. Recent suggestions to improve on this have been put forward\cite{T-parity} \cite{Sc}. It is unclear to me that they work without excessive complications of the simplest models or after properly taking into account  the full LEP2 constraints\footnote{The T-parity\cite{T-parity} advocated in the case of the "littlest" model requires an essential extension of the fermion sector, as noticed in ref. (\cite{low}). I suspect that the LEP2 constraints on the "simplest" model may be stronger than claimed in ref. (\cite{Sc}).}. With a scale $f$ in the  $1 \div 2$ TeV range, the mass of the heavy top might be in the same range, depending on the details of the model, with a consequent number of events useful for detection at the LHC\cite{PPP}.

\section{On the role of a Linear Collider}

Assuming a moderate fine tuning, every solution of the EWSB problem discussed here should give at the LHC some detectable signal, in spite of their variety in the different cases. This is indeed the strength of the Large Hadron Collider. After the discovery of any such signal, one would like to get a thorough understanding of the overall picture. Without forgetting the possible role of many diverse "low energy" experiments or of the same LHC, it is here that the LC could prove essential\cite{LCrapp}. Always for concreteness, I briefly comment on this role in the various cases that I have discussed and can be viewed in some way as benchmarks.

\subsection{ The MSSM or the NMSSM}

In a favorable point of the MSSM or of the NMSSM parameter space, the s-particles directly produced at a LC would allow a precise determination of the parameters themselves, which might in turn  elucidate some key properties of the underlying  high energy theory. If it occurs, this looks as the most promising impact of the LC. How favorable would the point in parameter space have to be for this to happen? Hard to tell  and, in any case, crucially dependent on the c.o.m. energy of the collider. The Higgs will in any event be measured with ease. In general, a part or even more than the mass, I think that the most interesting parameters are the top-Higgs coupling and the Higgs self coupling. To measure these parameters with high accuracy, at per-cent level, however, is challenging,  I understand. 

I have mentioned  that some extensions of the MSSM are designed to raise the Higgs mass significantly, even well above the 150 GeV limit of the NMSSM, as a way to alleviate the fine tuning problem. At some point this becomes problematic with the EWPT, - I have also recalled -, unless one adjusts some extra parameter. If taken seriously, this would be another reason to push up the c.o.m energy of the LC.

\subsection{ Supersymmetry broken by boundary conditions on an extra dimension}

The s-particle spectrum in this case has only the stops as relatively light fragments, which could however be as heavy as a TeV or so. What would a 500 GeV Linear Collider see in this case other than a light Higgs, in the $110 \div 125$ GeV range? There could be a second Higgs-like supermultiplet with a scalar component without a vev\cite{two-loops}. Otherwise the LC could reveal the low energy tale of some 4-fermion interaction mediated by the Kaluza Klein towers of the gauge bosons\footnote{Momentum conservation in the fifth dimension inhibits the coupling of these KK vectors to the light fermions\cite{BHN}. This conservation, however, is broken by kinetic terms localized on the boundaries of the extra dimension.}.  Quite in general at a LC this is a sensitive signal of new physics with a mass gap. Defining as $\Lambda$ in

\begin{equation}
\mathcal{L}=\frac{1}{\Lambda^2} (\bar{f} \gamma_{\mu} f)_{L,R}
(\bar{f} \gamma_{\mu} f)_{L,R}
\end{equation}
the effective scale characterizing these interactions, the sensitivity reach of a 500 GeV LC with polarized beams ranges from 12 to 25 TeV, depending on the fermion and on its helicity\cite{study-group}. This is to be compared, for example, with the current LEP2 limits which are in the $3 \div 7$ TeV range for the individual interactions.

\subsection{Split Supersymmetry}

A long lived gluino discovered at the LHC would be an indication of split supersymmetry, but not a proof of it. The Higgs mass could still not be fine tuned at all. To prove the contrary,  it would be crucial to produce the gauginos at a LC and precisely measure their couplings to the Higgs supermultiplets\cite{split-susy}. The naturalness argument does not act anymore as an upper bound on the masses on these weak gauginos. Rather it is the requirement of a consistent relic neutralino density that makes them lighter than $1 \div 2$ TeV or so\cite{split-susy}. The LC might have to be well above the TeV c.o.m. energy to produce and study them. The Higgs mass in the case of split supersymmetry is expected below 160 GeV or so.

\subsection{Little Higgs models}

In this case the signal to be looked for at a LC would again be a 4-fermion interaction mediated by the exchange of a heavy gauge boson. I have already mentioned the model dependence of their couplings to the light fermions. 

Quite in general, i.e. here as in many other possible cases, one can also consider an indirect signal on the electroweak precision observables. Precise measurements of the weak mixing angle, the W-mass and the top mass, all equally essential\footnote{I am not considering here a possible significant improvement in the determination of $\alpha(M_Z)$ since I do not know if it will happen or, in case, how to estimate it. Such an improvement might change the numbers in the following of this paragraph as the entire logic of the comparison between theory and experiment.}, would lead to a determination of the Higgs mass, in a given theory, e.g. the SM, to be compared with the direct measurement. The current relative uncertainty on the indirect Higgs-mass determination is about $60 \%$, with some uneasiness on the important input from $\sin^2 \theta_{eff}$ \cite{alt}. A 500 GeV LC might bring this uncertainty to $10 \div 15 \%$ level or even somewhat better by operating in the Giga-Z mode. Note however that the LHC is expected to go to a $15 \div 20 \%$ error and that the direct measurement of the Higgs mass should be by far more precise.

\section{Summary and conclusions}

In discussing our current understanding  of the EWSB problem, a need of concreteness suggests that we restrict our attention to "calculable" models, as defined in Sect. \ref{int}. If we do so, two physical principles emerge: Supersymmetry or the Higgs as a Pseudo-Goldstone Boson. Supersymmetry has far reaching consequences. The lack of signals so far, especially but not only in the Higgs search, stimulates a debate and ideas on its specific realization, which deserve attention. The PGB picture of the Higgs has also been around for quite some time. It is not easy to make it work up to the end in a natural way. It has nevertheless received new stimulus from the model building in extra dimensions, which have led in turn to the Little Higgs models in 4D. They involve tricks which look to me somewhat {\it ad hoc}. They work, however.

These and other ideas on the EWSB problem imply a variety of possible signals at the LHC. The great virtue of the LHC is that they should all be detectable, if naturalness is a good guide. To rely on naturalness, i.e. on a moderate fine tuning,  is not a logical necessity and it is not even free of ambiguities. It is convenient, though, and a common assumption in doing science, it appears to me. In the past of physics it has been successfully  applied in several important instances. 

If the LHC sees some signals of the EWSB mechanism, their full use and interpretation will almost inevitably require complementary experiments. The role of a Linear Collider would in particular be very significant in many of the cases that I have discussed, with its c.o.m. energy as a crucial parameter, hard to optimize, however, without further direct informations.

On general grounds, I would defend the following conclusions/comments:
\begin{itemize}
\item There are many reasons that make the EWSB the most (?) compelling/promising open problem in particle physics. 
\item Energies above the Fermi scale, i.e. the characteristic scale of EWSB, have only been scantily explored so far in a direct way.
\item A variety of options to address the EWSB problem (although not all on equal footing) have been put forward and deserve attention with an open mind.
\item Uncovering the mechanism of EWSB would definitely be a revolution in fundamental physics and would allow to put on a much firmer basis than it is possible now any further extrapolation for physics beyond the SM.
\end{itemize}

\section*{Acknowledgments}
I thank Toni Gherghetta, Riccardo Rattazzi, Alessandro Strumia, Alex Pomarol, Andrea Romanino, Lawrence Hall, Yasunori Nomura, Michele Papucci, Guido Marandella, Raman Sundrum for many useful discussions and comments.
This work is  supported  in part by MIUR and by the EU under TMR contract
HPRN-CT-2000-00148.

\end{document}